\documentclass[
 reprint,
superscriptaddress,
aps,
prl,
 amsmath,amssymb,
floatfix,
]{revtex4-1}
\usepackage{dcolumn}
\usepackage{bm}
\usepackage{geometry}
\usepackage{graphicx} 

\usepackage{float}
\usepackage{multirow}
\usepackage{booktabs} 
\usepackage{array} 
\usepackage{paralist} 
\usepackage{verbatim} 
\usepackage{subfig} 
\usepackage{rotating}
\usepackage{rotating}
\usepackage{tikz}	
\usetikzlibrary{shapes,snakes}
\usetikzlibrary{backgrounds,fit,decorations.pathreplacing}  

\usepackage{fancyhdr} 
\usepackage{threeparttablex}
\usepackage{amsmath, amsthm}
\usepackage{afterpage}
\usepackage[nottoc,notlof,notlot]{tocbibind} 
\usepackage[titles,subfigure]{tocloft} 

\graphicspath{{./figures/}}
\DeclareGraphicsExtensions{.eps}

\usepackage{caption}
\begin{document}

\title{Quantum Coherence and Entanglement in the Avian Compass}
\author{James A. Pauls}\affiliation{Goshen College, Goshen, IN 46526 USA}
\author{Yiteng~Zhang}\affiliation{Department of Physics, Purdue University, West Lafayette, IN, 47907 USA}
\author{Gennady P. Berman}\affiliation{Theoretical Division, T-4, MS B-213, Los Alamos National Laboratory, Los Alamos, NM 87545 USA}
\author{Sabre~Kais}\email{kais@purdue.edu}
\affiliation{Department of Chemistry, Department of Physics and Birck Nanotechnology Center, Purdue University,
West Lafayette, IN 47907 USA}
\affiliation{Qatar Environment and Energy Research Institute, Qatar Foundation, Doha, Qatar}
\begin{abstract}
The radical pair mechanism is one of two distinct mechanisms used to explain the navigation of birds in geomagnetic fields. However, little research has been done to explore the role of quantum entanglement in this mechanism. In this paper, we study the lifetime of radical pair entanglement corresponding to the magnitude and direction of magnetic fields to show that the entanglement lasts long enough in birds to be used for navigation. We also demonstrate that, due to a lack of orientational sensitivity of the entanglement in the geomagnetic field, the birds are not able to orient themselves by the mechanism based directly on radical-pair entanglement. To explore the entanglement mechanism further, we propose a model in which the hyperfine interactions are replaced by local magnetic fields of similar strength. The entanglement of the radical pair in this model lasts longer and displays an angular sensitivity in weak magnetic fields, both of these factors are not present in the previous models.
\end{abstract}
\maketitle

$Introduction.$---The ability of many animal species, such as birds, insects, and mammals to sense the geomagnetic field for the purpose of orientation and navigation has led to huge interest in the field of biophysics \cite{mag0,mag}. There are currently two leading hypotheses to explain this remarkable ability: the magnetite-based mechanism, and the radical pair mechanism \cite{Vedral,Schulten,Cai,cc1,WuDickman}. The latter mechanism has been supported by results in the field of spin chemistry \cite{spin1,spin2} and by biological experiments \cite{cc0}. Recently several authors have raised the intriguing possibility that living systems may use nontrivial quantum effects to optimize their orientation behavior \cite{Schulten,Vedral,nature,Ban}.

It has been suggested that entanglement, rather than mere quantum coherence is the contributing factor which allows the avian compass to achieve its high level of sensitivity \cite{Cai}. If this is so, does the duration of the entanglement last long enough to impact biological processes, and is the  entanglement sensitive enough to the inclination of the radical pair with respect to the Earth's magnetic field? To answer these questions, we examine the lifetime of radical pair entanglement corresponding to different magnetic field strengths, and compare the results with the candidate chemical reaction \cite{Schulten}. We also study the angular dependence of the radical pair entanglement within the geomagnetic field. Based on our results, we propose a new model to explore the underlying details. We find that the entanglement present in our proposed model displays both directional sensitivity as well as a sufficiently long duration of entanglement.

\begin{figure}[htbp]
\includegraphics[width=0.52\textwidth,height=0.27\textwidth]{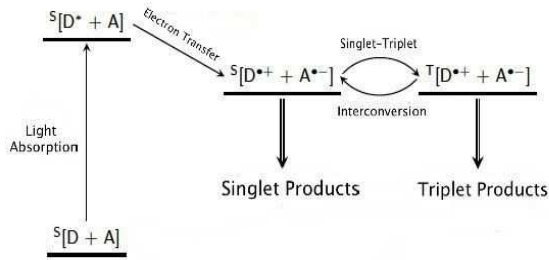}
\caption{{\small Scheme of RPM. After a light-induced electron excitation, the donor transfers an electron to the acceptor,  forming a radical pair with the acceptor molecule. The interconversion between singlet and triplet is affected by an external magnetic field. Finally the singlet and
triplet decay into different products.}}
\label{fig:figs}
\end{figure}

$Model.$---The basic scheme of the Radical Pair Mechanism (RPM) involves three steps \cite{Ritz,Tiersch}. The first step is light absorption, which is then followed by the formation of the radical pair and its interconversion between the singlet and triplet for electron spins. The final step is the decay of the singlet and triplet states to chemical products which produce a chemical signal detectable through an, as yet unknown, biological pathway. Typically, the radical-pair reaction involves two kinds of molecules which play the roles of electron donor and acceptor. After absorbing light, the electronic state of the donor molecule is excited, e.g. one of the electrons in the donor molecule is excited from the highest occupied molecular orbital (HOMO) to the lowest unoccupied molecular orbital (LUMO)  \cite{Tiersch}. In the following step, if the two molecules are close enough to each other, the donor electron will be transferred to the acceptor molecule. Thus, both molecules will contribute an unpaired electron to form a pair of radicals. These two unpaired electrons in the donor and acceptor molecules are initially bound in a singlet state before they are spatially separated. Under the effect of the external magnetic field, i.e. the geomagnetic field and local nuclear spins, the electron pair tranfers between the singlet and triplet states. Finally, the singlet and triplet states will produce different reaction products \cite{Ritz,Tiersch}. The whole process is shown in Fig. \ref{fig:figs}.

To investigate the role of entanglement in this chemical compass, we focus on the second step of the RPM scheme, since it is while the electrons are separated that it is believed the geomagnetic field may influence intersystem conversion.

Based on earlier literature \cite{Tiersch}, we include only the Zeeman interaction and the hyperfine interaction in the Hamiltonian of the system \cite{Ritz}:
\begin{equation}
H=g \mu_B \sum_{i=1}^2 \vec S_i \cdot \left( \vec B + \widehat A_i \cdot \vec I_i \right). \label{eq:eqs}
\end{equation}
In Eq. (\ref{eq:eqs}), the first term accounts for the Zeeman interaction, and the second term for the hyperfine interaction. (We assume that each electron is coupled to a single nucleus.) $\vec I_i$ is the nuclear spin operator; $\vec S_i$ is the electron spin operator, i.e., $\vec S={\vec\sigma}/{2}$ with $\vec\sigma$ being the Pauli matrices; $g$ is the $g$-factor of the electron, which is chosen to be $g=2$; $\mu_B$ is the Bohr magneton of the electron; and $\widehat A_i$ is the hyperfine coupling tensor, a 3$\times$3 matrix.

As suggested in Ref. \cite{Ritz}, we model the radical-pair dynamics with a Liouville equation,
\begin{align}
\dot{\rho}(t)=&-\frac{i}{\hbar}[H,\rho(t)] \nonumber\\
                    &-\frac{k_S}{2}\left\{Q^S,\rho(t)\right\}-\frac{k_T}{2}\left\{Q^T,\rho(t)\right\}. \label{eq2}
\end{align}
In Eq. (\ref{eq2}), $H$ is the Hamiltonian of the system; $Q^S$ is the singlet projection operator, i.e. $Q^S=|S\rangle\langle S|$, and $Q^T=|T_+\rangle\langle T_+|+|T_0\rangle\langle T_0|+|T_-\rangle\langle T_-|$ is the triplet projection operator, where $|S\rangle$ stands for the singlet state and ($|T_+\rangle, |T_0\rangle, |T_-\rangle$) stand for the triplet states\cite{Lambert}; $\rho(t)$ is the density matrix for the system; $k_S$ and $k_T$ are the decay rates for the singlet state and triplet states, respectively.

$Calculations~and~Results.$---For our calculations we assume that the initial state of the radical pair is a perfect singlet state, $\mid$S$\rangle=\frac{1}{\sqrt{2}}(\mid\uparrow\downarrow\rangle-\mid\downarrow\uparrow\rangle)$. Therefore, the initial condition for the density matrix is: $\rho(0)=\frac{1}{4}\hat{I}_N\otimes {Q}^S$, where the electron spins are in the singlet states, and nuclear spins are in a completely mixed state, which is a 4$\times$4 identity matrix. Assuming that the recombination rate is independent of spin, the decay rates for the singlet and triplet should be the same \cite{Ritz}, $k_S=k_T=k$, i.e., $k$ is the recombination rate for both the singlet and triplet states. The external weak magnetic field, $\vec{B}$, representing the Earth's magnetic field in Eq. (\ref{eq:eqs}), depends on the angles, $\theta$ and $\varphi$, with respect to the reference frame of the immobilized radical pair, i.e., $\vec{B}=B_{0}(\sin\theta\cos\varphi, \sin\theta\sin\varphi, \cos\theta)$, where $B_0=0.5$G is the magnitude of the local geomagnetic field. Without losing the essential physics, $\varphi$ can be assumed to be $0$.

Since the radical pair must be very sensitive to different alignments of the magnetic field, it is necessary to assume that the hyperfine coupling tensors in Eq. (\ref{eq:eqs}) are anisotropic. However, for the sake of simplicity, we employ the hyperfine coupling as anisotropic for one radical, and the other as isotropic \cite{Ritz}, i.e.,

\begin{displaymath}
\widehat{A_1}=
\left(\begin{array}{ccc}
$10G$ & $0$ & $0$\\
$0$      &$10G$ &$0$ \\
$0$& $0$& $0$
\end{array}\right)
\normalsize{,}
\
\widehat{A_2}=
\left(\begin{array}{ccc}
$5G$ & $0$ & $0$\\
$0$      &$5G$ &$0$ \\
$0$& $0$& $5G$
\end{array}\right)
\end{displaymath}
\begin{figure}[htbp]

\includegraphics[width=0.5\textwidth,height=0.4\textwidth]{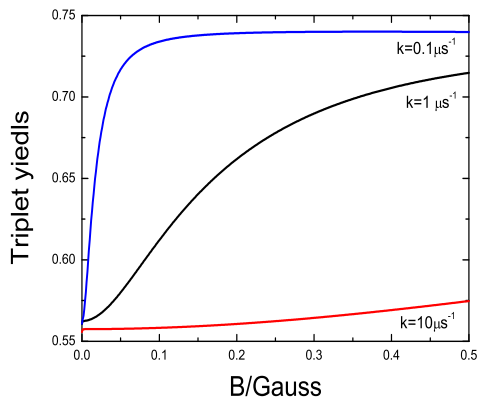}
\caption
{\small{The triplet yields for three different decay rates as a function of the external magnetic field magnitude. The black line, for which $k=1\mu$s$^{-1}$ seems to be a reasonable curve.}}
\label{decay}
\end{figure}

In order to determine what values of the decay rates are reasonable for biological systems, we calculate the influence of different decay rates on the triplet yield, $\Phi_T$, as the external magnetic field varies, by using the parameters defined above.   We define the triplet yield as \cite{Ban} \cite{Lambert}: $\Phi_T=k\int_0^\infty Tr[Q^T\rho(t)] dt$, where $Q^T=|T\rangle\langle T|$, and $\mid$T$\rangle=|T_+\rangle + | T_0\rangle+ | T_-\rangle$.
The effect of the radical pair decay rates on the triplet yield has a twofold function \cite{Ritz}. For a very high decay rate, i.e., larger than $10\mu$s$^{-1}$, the rapid decay of the radical pair prevents efficient singlet-triplet mixing, as can be seen by the increase of the triplet yield in the weak magnetic field. This means that the weak magnetic field has very little effect on the triplet yields with fast decay rates. However, for very slow decay rates, i.e., smaller than $0.1\mu$s$^{-1}$, the triplet yield increases up to its maximum almost immediately when the magnetic field increases from zero, but remains essentially static as the magnetic field continues to increase. A decay rate of the order of $1\mu$s$^{-1}$, seems to be optimum for the detection of a weak magnetic field. For all further calculations with this model we have assumed this value for our decay rate, i.e. $k=1 \mu$s$^{-1}$.

\begin{figure}[htbp]
\includegraphics[width=0.5\textwidth,height=0.4\textwidth]{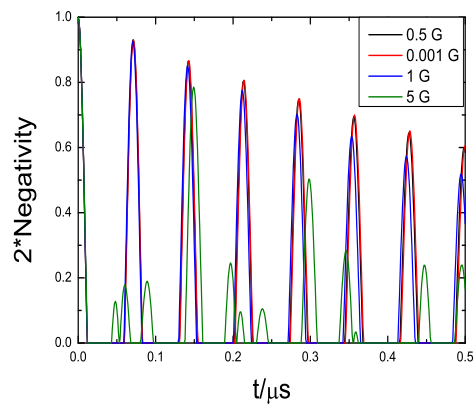}
\caption{\small{Entanglements for different magnitudes of the magnetic field for the angle $68^o$ between the $z$ axis of the radical pair and the magnetic field.}}
\label{entanglement}
\end{figure}

\begin{figure}[htbp]
\includegraphics[width=0.5\textwidth,height=0.4\textwidth]{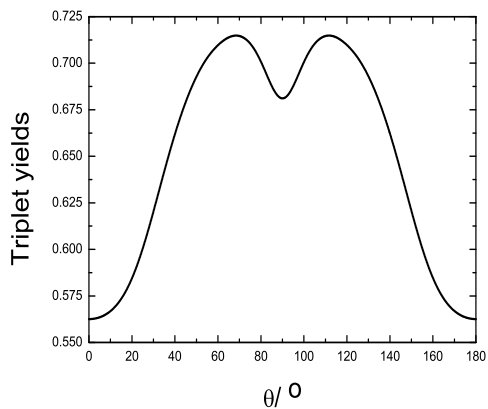}
\caption{\small {Angular dependence of the triplet yields. The triplet yields are symmetric about $90^o$. }}
\label{yields}
\end{figure}

Having fixed the decay rate to be $1\mu$s$^{-1}$, we study the radical pair entanglement as a function of the magnitude of the geomagnetic field. The $z$ axis of the radical pair is aligned at an angle of $68^o$ with the magnetic field vector, which is the angle at which an earth-strength magnetic field produces the largest triplet yield \cite{Ritz}. (See Fig. \ref{yields} in \cite{Ritz}.) In this paper, we use negativity as the metric of entanglement, $N(\rho)=\frac{\| \rho^{T_A} \|_1-1}{2}$, where $\| \rho^{T_A} \|_1$ is the trace norm of the partial transpose of the system's density matrix  \cite{neg}. The results are shown in Fig. \ref{entanglement}. We can see that when the magnetic fields are weaker than the Earth's magnetic field, or as strong as 1G, the  entanglement curves are almost identical. There does not appear to be any unique behavior that distinguishes a field in the neighborhood of 0.5 Gauss. However, under the Earth's magnetic field, the entanglement will be robust periodically during the first 0.5$\mu$s, which is longer than the suggested duration of radical pair separation \cite{Schulten}. A stronger magnetic field (e.g. 5G) will disturb this periodicity. Previous research on the magnetic-field sensitivity of the chemical compass has demonstrated that the entanglement is helpful only if nature allows birds to optimize this behavior \cite{Cai}. On these grounds one can say that the entanglement lasts long enough to play crucial role in the orientation of birds.

So, we can say that the entanglement could play a role in the orientation and navigation of birds. We now recheck one of the properties of the avian compass, that it should depend on inclination but not polarity. In Fig. \ref{yields}, we see that the triplet yields are symmetric about $90^o$. Consequently, the radical pair mechanism cannot distinguish between magnetic fields that are oppositely directed but have the same magnitude \cite{Ritz}.

\begin{figure}[htbp]
\includegraphics[width=0.5\textwidth,height=0.4\textwidth]{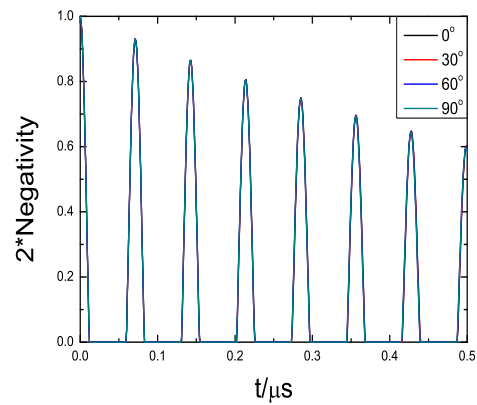}
\caption{\small{Entanglements for different angles. All curves are practically identical. In the geomagnetic field, entanglement does not change with orientation.}}
\label{angle}
\end{figure}

The surprising result, in Fig. \ref{angle}, is that the dynamics of entanglement does not change with angle, i.e., entanglement is not sensitive to the angle between the $z$-axis of the radical pair and the Earth's magnetic field. Therefore, it is reasonable to conclude that the entanglement of the radical pair cannot provide the same information as the triplet yields. In other words, using this model, the entanglement of the system does not directly affect the birds' ability to orient themselves. However, there might be indirect mechanisms which allow the birds to utilize entanglement.

The above result (Fig. \ref{angle}) shows that the dynamics of entanglement are nearly static for different angles under the symmetric hyperfine tensors. This raises the question, what will happen when we use an asymmetric hyperfine tensor? We examined several such cases. The asymmetric hyperfine tensors we applied are,
\begin{displaymath}
\widehat{A{^b_1}}=
\left(\begin{array}{ccc}
$10G$ & $0$ & $0$\\
$0$      &$10G$ &$0$ \\
$0$& $0$& $4G$
\end{array}\right)
\normalsize{,}
\
\widehat{A{^b_2}}=
\left(\begin{array}{ccc}
$5G$ & $5G$ & $0$\\
$0$      &$5G$ &$0$ \\
$0$& $0$& $5G$
\end{array}\right)
\end{displaymath}
and
\begin{displaymath}
\widehat{A{^c_1}}=
\left(\begin{array}{ccc}
$0$ & $0$ & $0$\\
$0$      &$0$ &$0$ \\
$0$& $0$& $4G$
\end{array}\right)
\normalsize{,}
\
\widehat{A{^c_2}}=
\left(\begin{array}{ccc}
$0$ & $5G$ & $0$\\
$0$      &$0$ &$0$ \\
$0$& $0$& $0$
\end{array}\right)
\end{displaymath}

\begin{figure}[htbp]
\includegraphics[width=0.5\textwidth,height=0.4\textwidth]{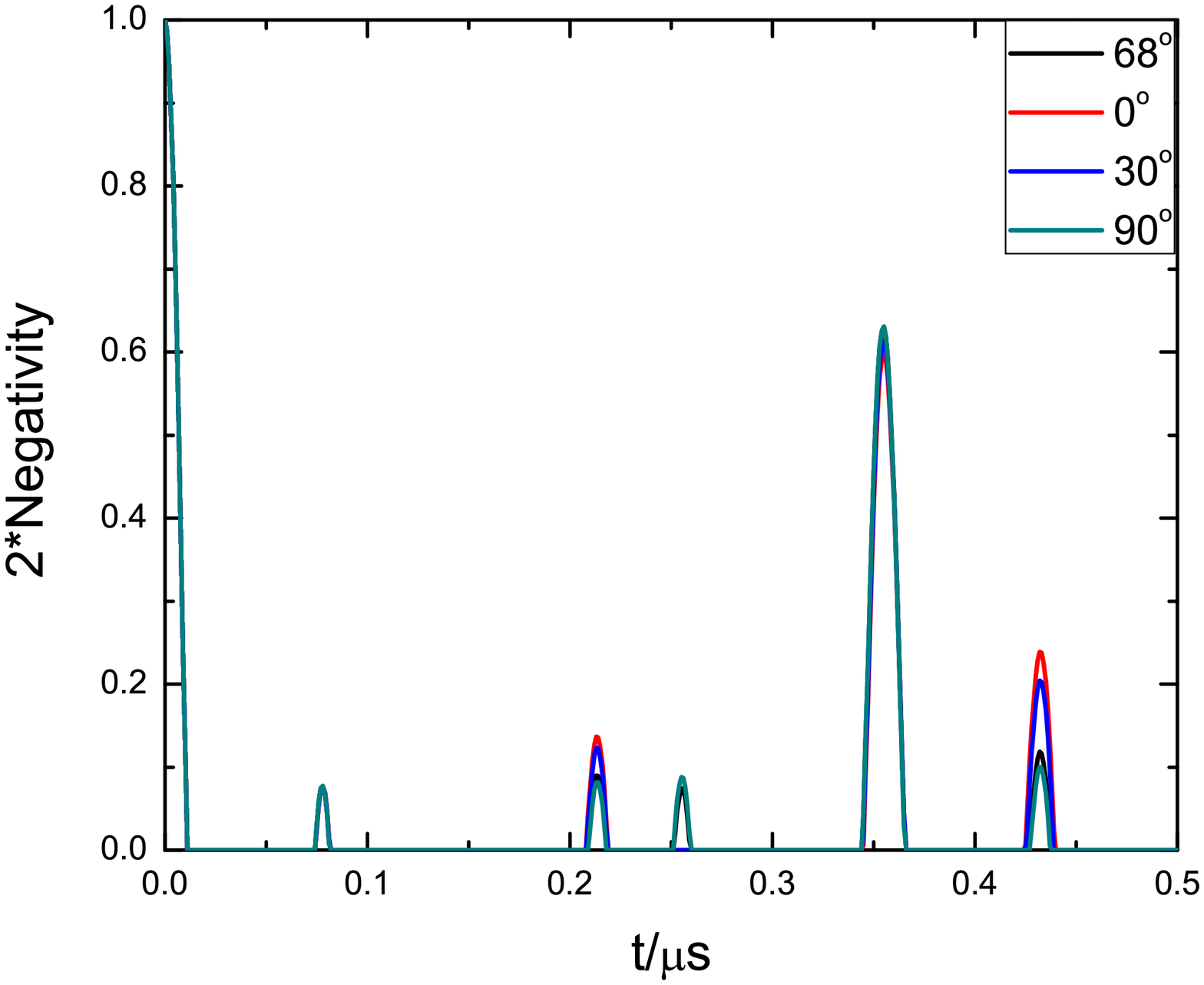}
\caption{\small{Entanglements for different angles under the hyperfine coupling tensors $\widehat A{^b_i}$.}}
\label{angle_1}
\end{figure}

\begin{figure}[htbp]
\includegraphics[width=0.5\textwidth,height=0.4\textwidth]{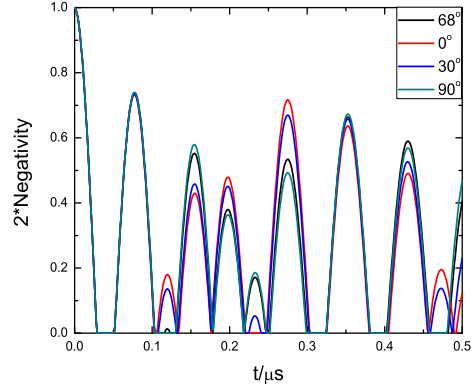}
\caption{\small{Entanglements for different angles under the hyperfine coupling tensors $\widehat A{^c_i}$.}}
\label{angle_2}
\end{figure}

From Fig. \ref{angle_1} and Fig. \ref{angle_2}, we can easily conclude that the hyperfine coupling tensor pair of $\widehat A{^c_i}$ gives an intriguing result, the dynamics of the entanglement is clearly dependent on the system's orientation. This result inspired us to develope a new model in which only the external magnetic fields are considered, as in the case of $\widehat A{^c_i}$ there are only two non-zero terms totally.

$New~Model.$--- Previously, we had assumed that one electron of the radical pair experiences an anisotropic hyperfine coupling, while the other experiences an isotropic one. However, this model cannot produce an angular-sensitive entanglement. On the other hand, the hyperfine coupling tensors $\widehat A{^c_i}$ led to an angular-sensitive result. Inspired by this result, we propose that each electron interacts with additional local magnetic fields, $\vec B_i$, rather than with the hyperfine fields. The Hamiltonian for this model is given by Eq. (\ref{eq:eqs}), but with $\widehat A_i$$\cdot$ $\vec I_i$ replaced by $\vec B_{i}$, the local magnetic field for the $i^{th}$ electron spin. We take the local fields to be, $\vec B_1=(0,0,4G)$, $\vec B_2=(0,5G,0)$.

In this case, we use the violation of the CHSH (Clauser, Home, Shimony, and Holt) inequality \cite{CHSH} as a witness of entanglement, a version of Bell's inequality \cite{Bell}. The CSHS inequality is given by $|E|\leq2\lambda_{max}^2$, where $|E|=\mid E(0,0)+E(0,t)+E(t,0)-E(t,t)\mid$, $E(t_1,t_2)=\langle \phi_{t_1} \mid(\vec \sigma_1 \cdot \vec a)(\vec \sigma_2 \cdot \vec b) \mid \phi_{t_2} \rangle$ is the two-time correlation function for a spin pair, and $\vec a$ and $\vec b$ are the unit direction vectors. The quantity, $\lambda_{max}$, is the maximum eigenvalue for the measurement operator, $(\vec \sigma_1 \cdot \vec a)(\vec \sigma_2 \cdot \vec b)$, which for our specific operator is equal to 1. When $|E|$ exceeds $2\lambda_{max}^2=2$, the correlation between the two spins can no longer be explained classically, so the system is entangled.

\begin{figure}[]
\includegraphics[width=0.22\textwidth,height=0.23\textwidth]{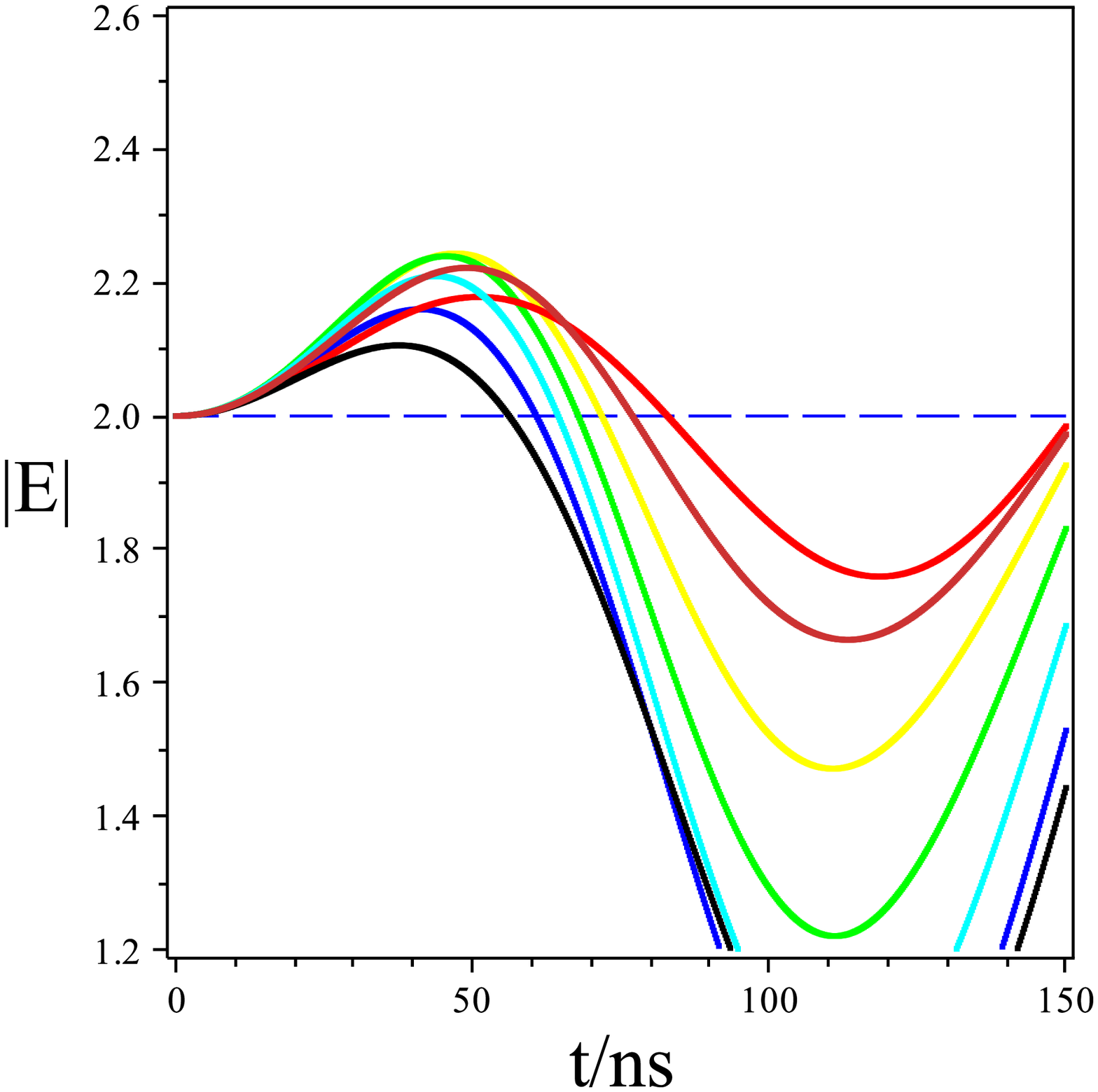}
\includegraphics[width=0.22\textwidth,height=0.23\textwidth]{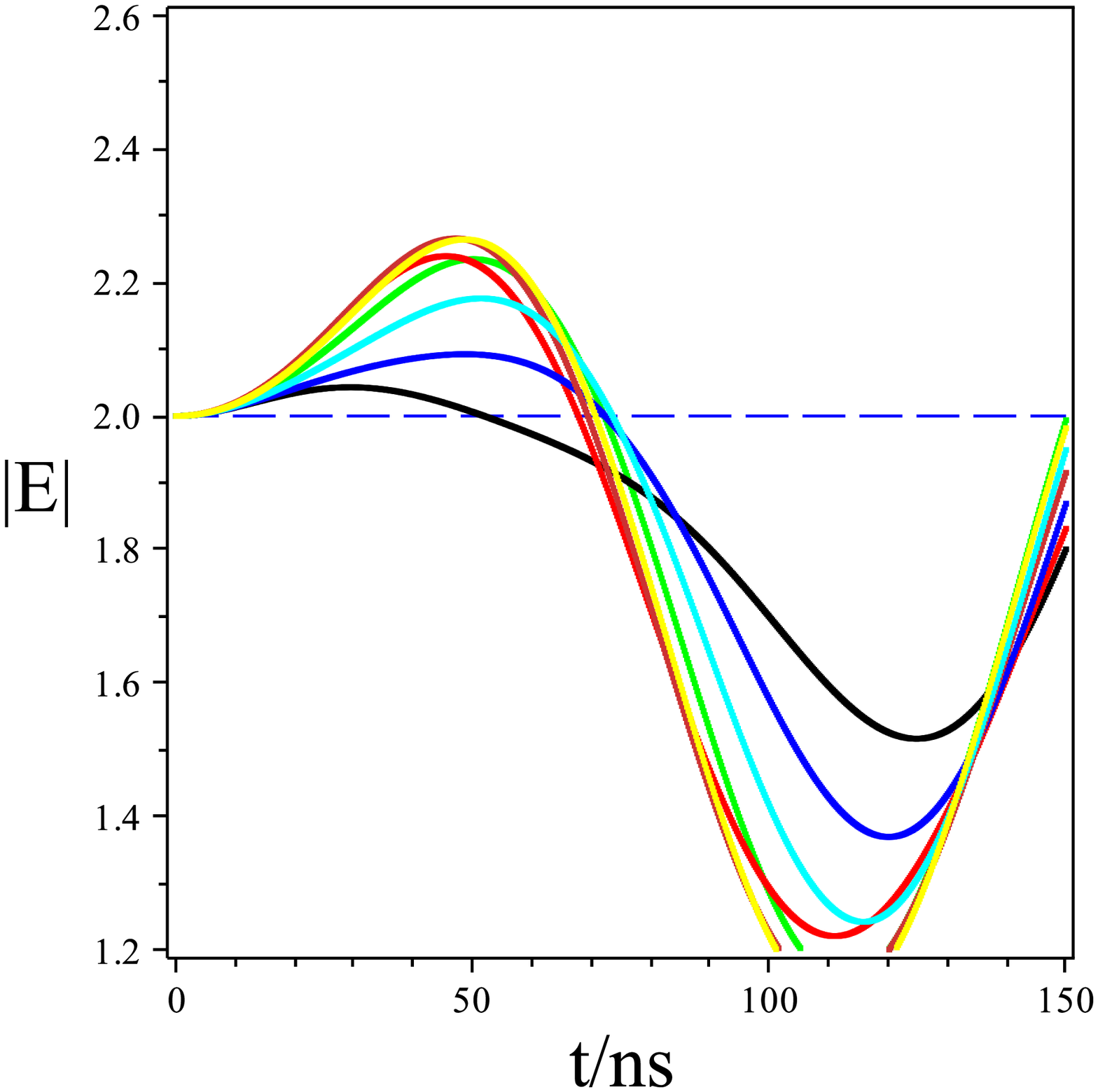}
\includegraphics[width=0.22\textwidth,height=0.23\textwidth]{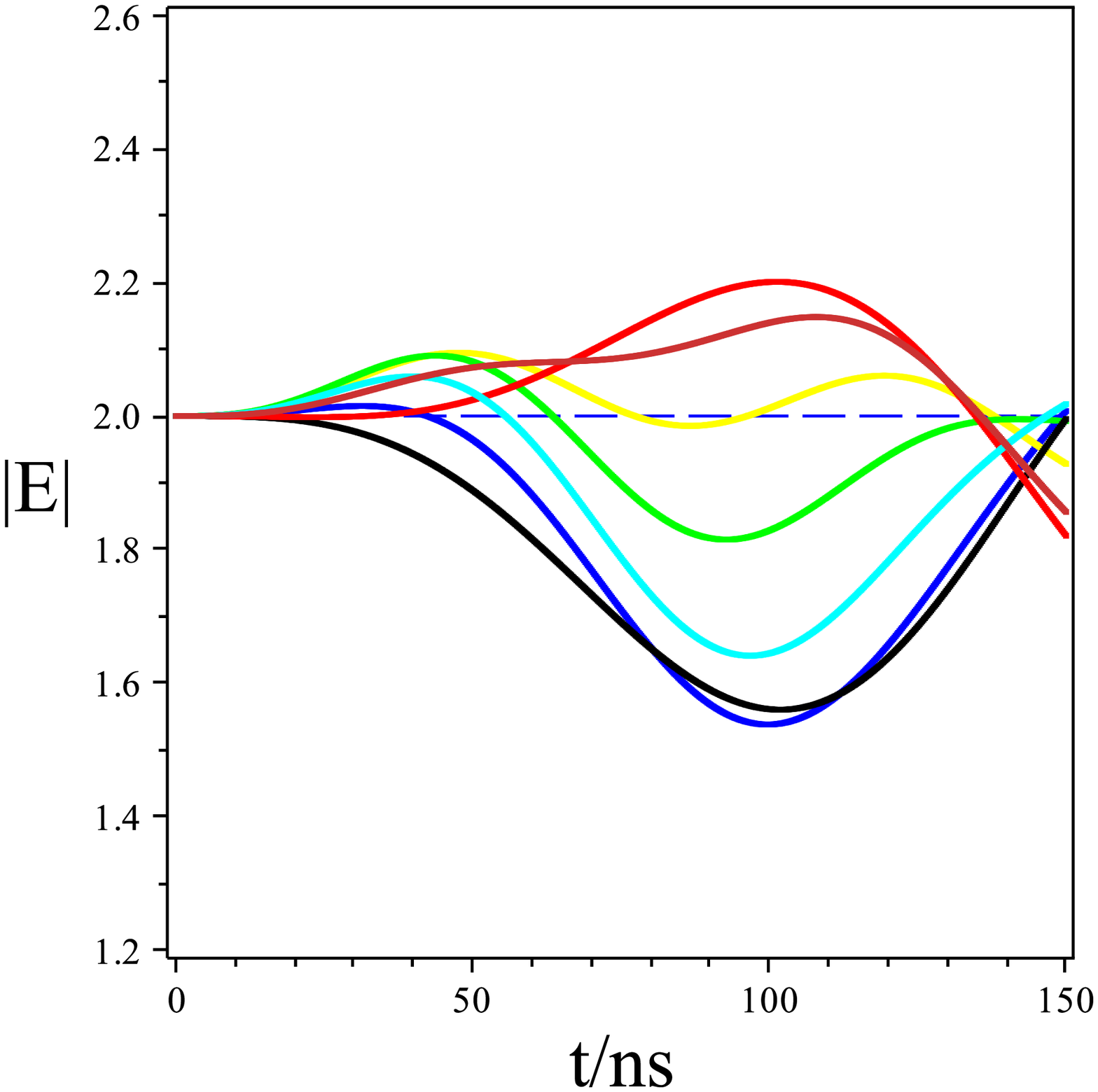}
\includegraphics[width=0.22\textwidth,height=0.23\textwidth]{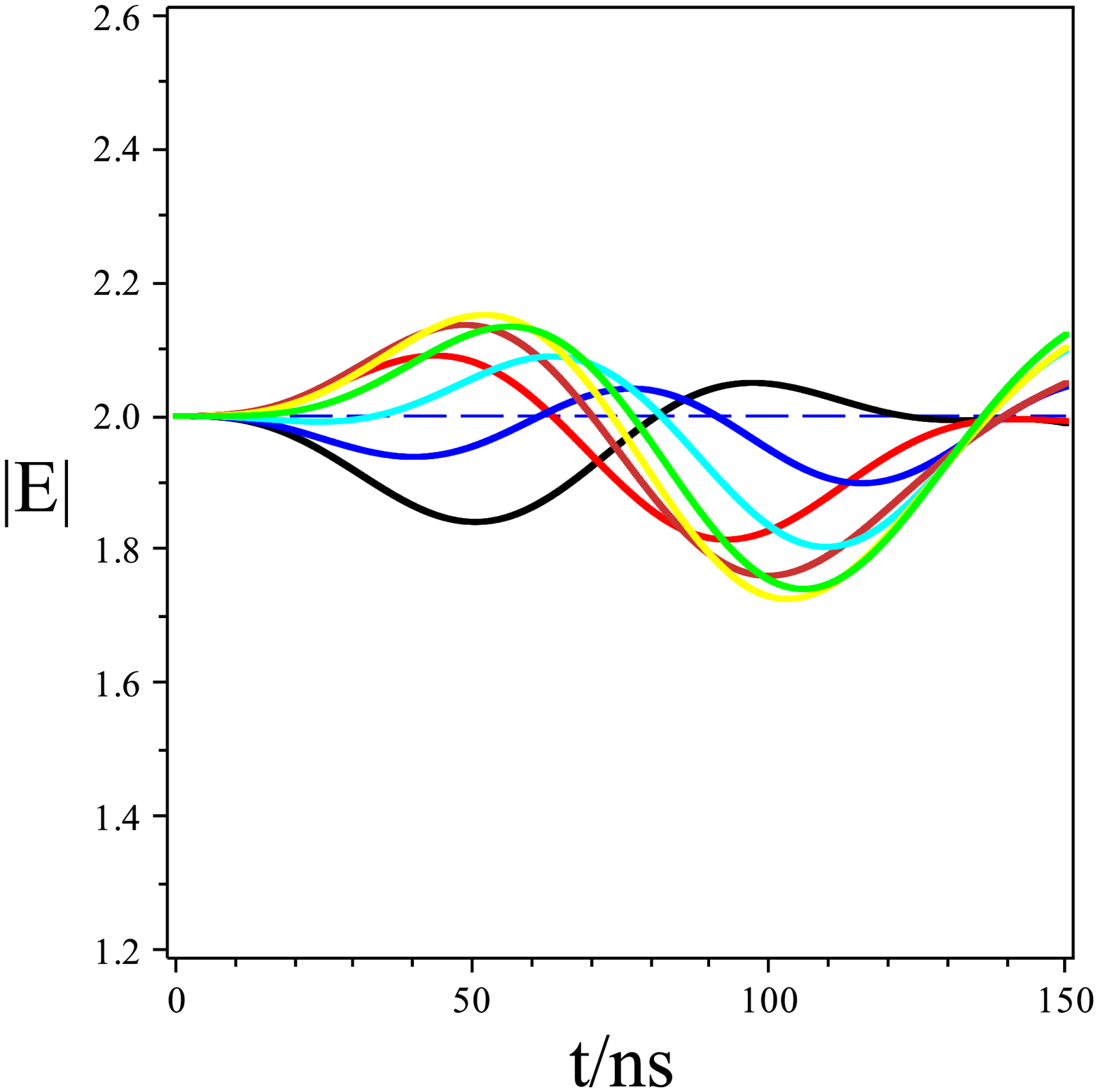}
\caption{\small {Polar and azimuthal dependence of the CHSH inequality. The dashed blue line represents the points above which the system is entangled. The various orientations are given by the lines: red ($0^o$), orange ($30^o$), yellow ($60^o$), green ($90^o$), light blue ($120^o$), dark blue ($150^o$), black ($180^o$). For the upper two figures, $\vec B_1=(0,0,5G)$ and $\vec B_2=(0,5G,0)$. The upper left figure depicts the azimuthal dependence for fixed $\theta=0^o$, while the upper right figure depicts polar dependence for fixed $\varphi=90^o$. For the lower two figures, $\vec B_1=(0,0,5G)$ and $\vec B_2=(0,5G,0)$. The bottom left and bottom right figures similarly depict the azimuthal and polar dependency, respectively. For all four figures $k=1\mu$s$^{-1}$.}}
\label{newmodel}
\end{figure}

Fig. \ref{newmodel} shows the CHSH inequality as a function of time for various orientations of the system in a magnetic field of 0.5G. Because there are now two perpendicular fields acting on the system, it becomes necessary to consider azimuthal orientation in addition to polar orientation. As seen in Fig. \ref{newmodel}, as $\theta$ increases from $0^o$ to $180^o$, the time for which the electron pair is entangled increases from roughly 60 ns to nearly 90 ns, while for $\phi$ from  $0^o$ to $150^o$ the variation of time of entanglement is restricted to an interval of less than 10 ns. It is interesting to note that this variation in time of entanglement occurs roughly on the same 100 ns time scale that the two electrons remain separated \cite{Schulten}.

Changing the relative angles and strengths of the local magnetic fields has a dramatic impact on the angular sensitivity. A change in the field strength of the first electron from 4G to 5G is enough to dramatically increase both the azimuthal and angular sensitivity of the entanglement.

If indeed a protein such as cryptochrome is in part responsible for magnetoreception, there must be some directional bias of the orientation of the protein, so that there will be a strong net signal. It is possible that this directional dependence could be provided by embedding within the membrane shelves of the photoreceptor cells. This form of embedding leaves the protein free to rotate about one axis, but greatly restricts the rotation about its second axis \cite{membrane}. For this reason it is necessary for the RP compass to be sensitive to rotation about one axis, while being virtually unaffected by rotation about the second. If the RP compass were to be sensitive to rotation in both $\theta$ and $\phi$, the result of randomly oriented proteins about the $\theta$ axis would average out to create a background signal that could potentially reduce the contrast of the RP compass.

At the present time little is known about how cryptochrome is situated within the retina, in particular how it embeds onto or within the cell membrane \cite{membrane}. There is no reason to assume that the z-axis of the RP model coincides with the fixed rotational axis of the embedded protein. As such, a configuration such as $\vec B_1=(0,0,5G)$ and $\vec B_2=(0,5G,0)$ might still produce a strong directional response under a coordinate transformation to the axis of protein rotation.

It should be pointed out that this model, unlike the previous model and its variants, is not symmetric about $90^o$, but  it is symmetric about $180^o$. While this might seem to contradict an inclination-only compass model, it is reasonable to assume that cryptochrome is either bound to both sides of the cell membrane, or embedded within the membrane in both up and down orientations, so that the net signal cannot discern the polarity of the geomagnetic field.

$Conclusions~and~Future~Work.$---We have identified that the entanglement decay rate is one of the key factors in the radical pair mechanism from the change of triplet yields (Fig. \ref{yields}). Also, we confirmed that the entanglement endures long enough for living systems to conduct the entanglement-based reactions. However, the dynamics of the entanglement is not sensitive to the change of angle between the $z$ axis of the radical pair and the geomagnetic field vector in the hyperfine model. Therefore, if we still believe that entanglement plays a crucial role in the orientation of birds as demonstrated before, there must be an indirect mechanisms by which the entanglement can affect the birds' behavior.

In the future, we will adjust the decay rates, for example, using different values for the decay rates of the singlet state and the triplet state to improve our model. We will also attempt to find the hidden bridge between the entanglement of the radical pair and the determination of orientation in a magnetic field.

$Acknowledgements.$
We would like to thank the NSF Center for Quantum
Information  for Quantum Chemistry (QIQC), Award
No. CHE-1037992, for financial support.

The work by G.P.B. was carried out under the auspices of
the National Nuclear Security Administration of the U.S. Department of Energy
at Los Alamos National Laboratory under Contract No. DE-AC52-06NA25396.

\end{document}